\documentclass[%
 reprint,
 amsmath,amssymb,
 aps,
]{revtex4-1}

\usepackage{graphicx}
\usepackage{dcolumn}
\usepackage{bm}

\begin{document}

\preprint{APS/123-QED}

\title{Direct laser printing of chiral plasmonic nanojets by vortex beams}

\author{S. Syubaev${}^1{}^,{}^2$, A. Zhizhchenko${}^1$, A. Porfirev${}^3$, E. Pustovalov${}^1$, O. Vitrik${}^1{}^,{}^2$, Yu. Kulchin${}^2$, S. Khonina${}^3$, S. Kudryashov${}^4{}^,{}^5$, A. Kuchmizhak${}^1{}^,{}^2{}^,$}
  \email{alex.iacp.dvo@mail.ru}
\affiliation{${}^1$School of Natural Sciences, Far Eastern Federal University, 6 Sukhanova Str., 690041 Vladivostok, Russia\\
${}^2$Institute for Automation and Control Processes FEB RAS, 5 Radio Str., 690041 Vladivostok, Russia\\
${}^3$Samara National Research University, Moskovskoe shosse, 34, Samara 443086, Russia\\
${}^4$Lebedev Physical Institute, Leninskiy prospect 53, Moscow 119991, Russia\\
${}^5$ITMO University, St.-Petersburg 197101, Russia}

\begin{abstract}
Donut-shaped laser radiation, carrying orbital angular momentum, namely optical vortex, recently was shown to provide vectorial mass transfer, twisting transiently molten material and producing chiral micro-scale structures on surfaces of different bulk materials upon their resolidification. In this paper, we show for the first time that nanosecond laser vortices can produce chiral nanoneedles (nanojets) of variable size on thin films of such plasmonic materials, as silver and gold films, covering thermally insulating substrates. Main geometric parameters of the produced chiral nanojets, such as height and aspect ratio, were shown to be tunable in a wide range by varying metal film thickness, supporting substrates, and the optical size of the vortex beam. Donut-shaped vortex nanosecond laser pulses, carrying two vortices with opposite handedness, were demonstrated to produce two chiral nanojets twisted in opposite directions. The results provide new important insights into fundamental physics of the vectorial laser-beam assisted mass transfer in metal films and demonstrate the great potential of this technique for fast easy-to-implement fabrication of chiral plasmonic nanostructures.

\vspace{0.1 cm}
\noindent \textbf{OCIS codes:} (050.4865) Optical vortices; (140.3390) Laser materials processing; (220.4241) Nanostructure fabrication; (050.1940) Diffraction; (140.3300) Laser beam shaping
\end{abstract}

\pacs{Valid PACS appear here}
\maketitle


\section{\label{sec:level1}Introduction}
Chirality is a specific inherent feature, readily occurring in natural living systems at almost all length scales from snails and sea shells to chiral molecules and DNA. Artificially designed nanoscale chiral-shaped structures, mimicking their natural analogues owing to unique ways of interaction with optical radiation, possess remarkable properties as circular dichroism, enhancement of nonlinear signals, highly directional emission and photoactivity [1-4]. Meanwhile, utilization of state-of-the-art direct nanofabrication techniques based on ion- or electron beam milling for chiral nanostructures fabrication is rather challenging, which triggers, in its turn, search for novel pathways to produce such unique nanostructures. At nanometer scale, fabrication of chiral nanostructures can be provided by assembly of specifically designed chiral molecules via various self-assembling processes [5-8]. However, the number of the available materials as well as the achievable size of the produced structures are both limited.\\
\indent Recently, in a number of papers an alternative pioneering approach, employing optical radiation with specially designed intensity-, polarization- or phase states, was proposed for fabrication of chiral structures [9-19]. In particular, using inherent chirality of nanoparticles, synthesis of twisted nanoribbons through self-assembly in water solution under irradiation with continuous-wave circularly polarized laser radiation was demonstrated [9]. More importantly, vortex laser pulses, carrying simultaneously orbital (OAM) and spin (SAM) angular momenta, were demonstrated to twist transiently molten metal producing chiral-shaped micron-size needles [10,11]. Later, similar effects were used to produce chiral structures on the surface polymers [12-14] and semiconductors [15,16]. Based on their experiments with ablation of bulk tantalum targets [17], the authors revealed that mass transfer and handedness of the produced surface structures are associated with corresponding phase helicity (or OAM), while the polarization helicity (or SAM) accelerates/decelerates the movement of the molten material.\\
\begin{figure*}
\includegraphics[width=0.75\textwidth]{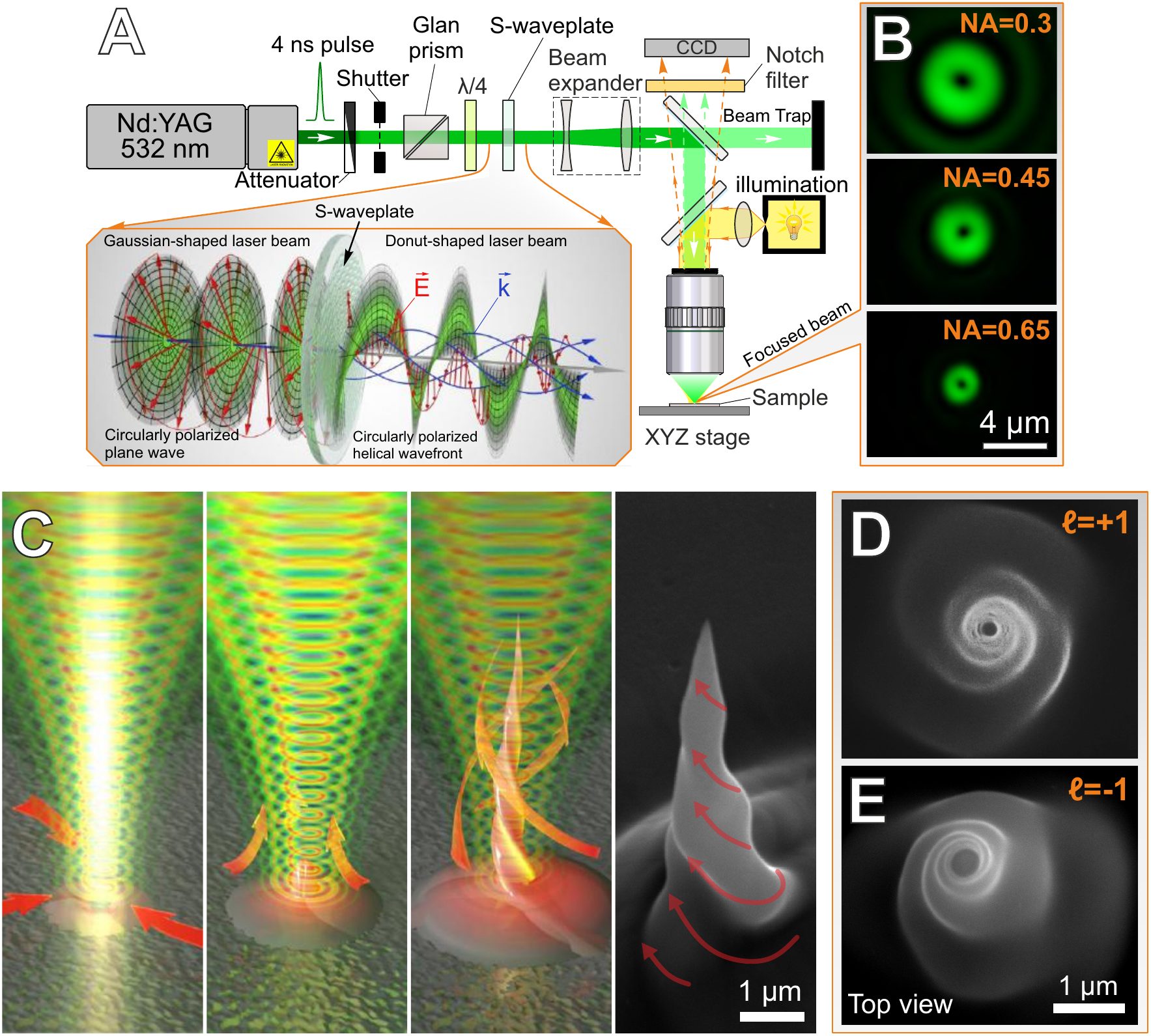}
\caption{(A) Schematic of the experimental setup for nanostructuring with the nanosecond vortex beams. (B) Vortex-beam intensity distributions measured in the focal plane of the two microscope objectives with NA=0.3 (left) and NA=0.65 (right). (C) Sketch schematically showing the formation process of twisted Ag nanojets under vortex pulse ablation. Right-most side-view SEM image shows twisted nanojet with the red arrows indicating the rotation direction. (D,E) Normal-view SEM images of twisted silver nanojets produced under single-pulse ablation of 500-nm thick Ag film with ns vortex pulses having opposite signs of vortex helicity.}
\end{figure*}
\indent Despite the apparent huge potential of such direct easy-to-implement method in fabrication of different chiral nanostructures mentioned in the previous studies, to the best of our knowledge, no experiments showing formation of twisted nanoneedles of variable size on surface of such common plasmonic materials, as silver or gold, were demonstrated so far. In this paper, we study formation of twisted nanojets under direct ablation of silver (Ag) films of variable thickness with nanosecond (ns) laser vortex pulses generated by passing circularly polarized radiation through S-waveplate [20]. Main geometric parameters of the produced chiral nanoneedles - height, width and aspect ratio - were shown to be tunable in a wide range by varying the thickness of the irradiated metal film, supporting substrate type, as well as optical size of the vortex beam. Moreover, nanosecond vortex pulse with donut-shaped intensity distributions, carrying two vortices with opposite handedness, was demonstrated to produce two chiral nanojets twisted in opposite directions. We believe the obtained results provide useful and important insights into fundamental picture of the vectorial-beam driven mass transfer in metal films, as well as demonstrate great potential of the direct vortex-beam ablation as a versatile fabrication technique for chiral nanophotonics and plasmonics.

\begin{figure*}
\includegraphics[width=0.75\textwidth]{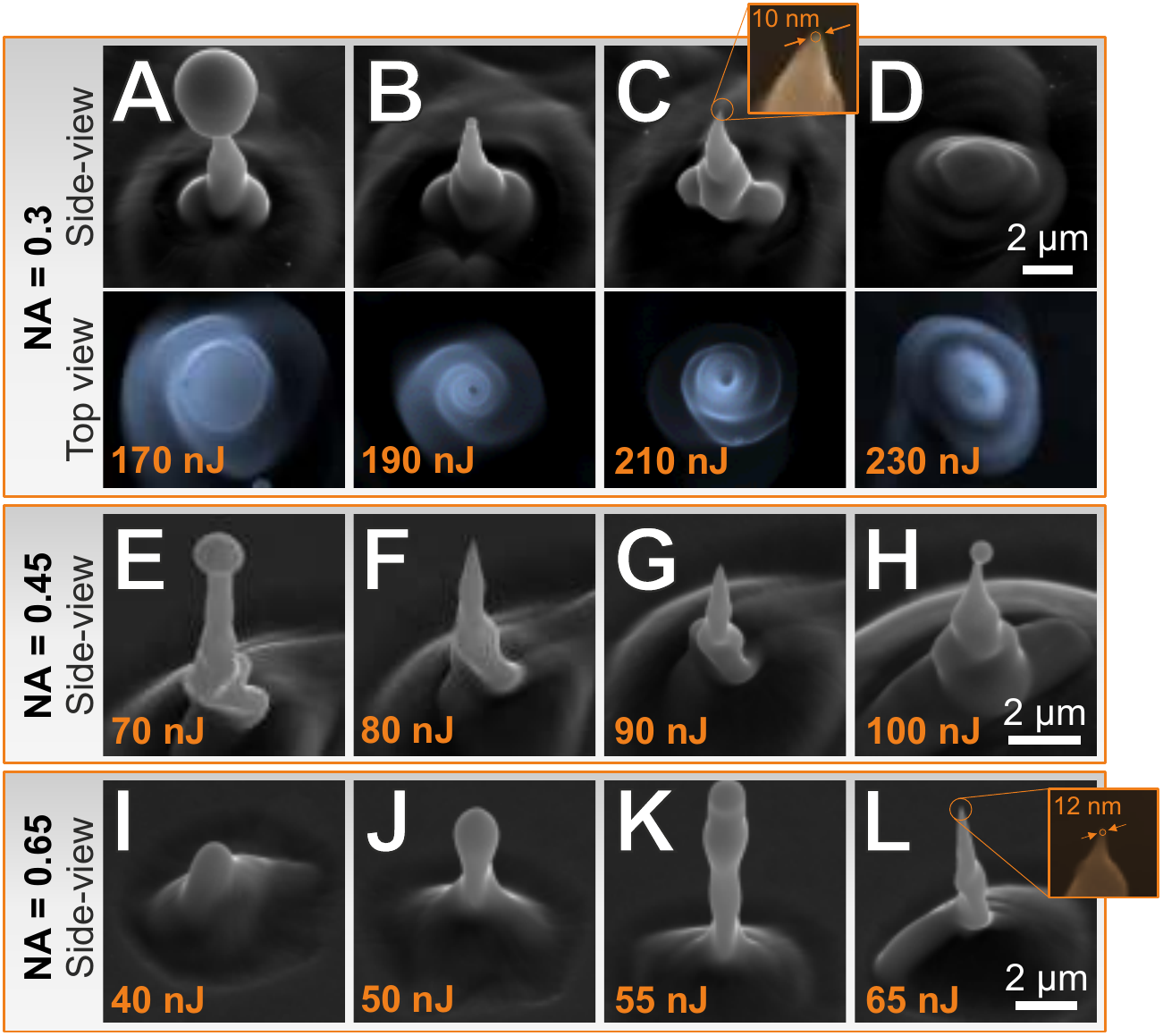}
\caption{(A-D) Side-view (view angle of 40°) and normal-view (in false color) SEM image pairs of twisted silver nanojets produced under single-pulse ablation of 500-nm thick Ag film with the vortex pulses focused at NA=0.3. Similar side-view (view angle of 40°) SEM images showing twisted nanojets fabricated on the surface of the same Ag film with single vortices focused with microscope objective at NA=0.45 (E-H) and NA=0.65 (I-L). The pulse energy in each image row increases from left to right. Insets in figures (C) and (L) show the magnified view of the nanojet tip indicating the typical curvature radii smaller than 12 nm.}
\end{figure*}

\section{\label{sec:level2}Experimental details}
A vortex beam, carrying OAM $\ell$=$\pm$1 (per photon, in units of $\hbar$), was generated by transferring second-harmonic radiation from a Nd:YAG laser system (Brio, Quantel: central wavelength – 532 nm, pulse duration – 8 ns, repetition rate – 20 Hz, maximal pulse energy – 50 mJ) through a Glan-Taylor polarizer and a quarter-waveplate (Fig. 1(A)), to produce a circularly-polarized beam, and finally through a commercially available radial polarization converter (S-waveplate, Altechna). Then, generated ns-laser vortex pulses were focused onto the sample surface by means of different microscope objectives (Xirox: NA = 0.45, 50x; Nikon Plan Fluor: NA = 0.3, 15x and NA = 0.65, 50x), yielding in donut-shape intensity distributions (Fig. 1(B)) with the outer optical diameter Dvortex of ~2.05 μm, ~2.7 μm and ~3.9 μm for NA=0.65, 0.45 and 0.3, respectively. As samples for nanostructuring, Ag films of variable thickness, ranging from 100 to 1000 nm, were used, being deposited by e-beam evaporation  (Ferrotec EV M-6) at the pressure of 5•10−6 bar and the average deposition rate of 0.5 nm/s onto optically smooth silica glass substrates. The resulting film thicknesses were pre-controlled by a calibrated quartz-crystal microbalance system (Sycon STC-2002), and then were verified by atomic force microscopy. Moreover, in order to study a substrate effect on formation of twisted nanojets, two types of substrates for 100-nm thick Ag films – silica glass and poly(methyl methacrylate) – were used. The samples were arranged on a PC-driven micropositioning platform (Newport XM and GTS series), providing a minimal translation step of 50 nm along each axis. Pulse energy E was varied by means of a transmission filter and controlled by a pyroelectric photodetector. All ablation experiments were performed under ambient conditions in the single-pulse mode. The morphology of the produced nanostructures was characterized by high-resolution scanning electron microscopy (SEM, Carl Zeiss Ultra 55+).
\begin{figure*}
\includegraphics[width=0.75\textwidth]{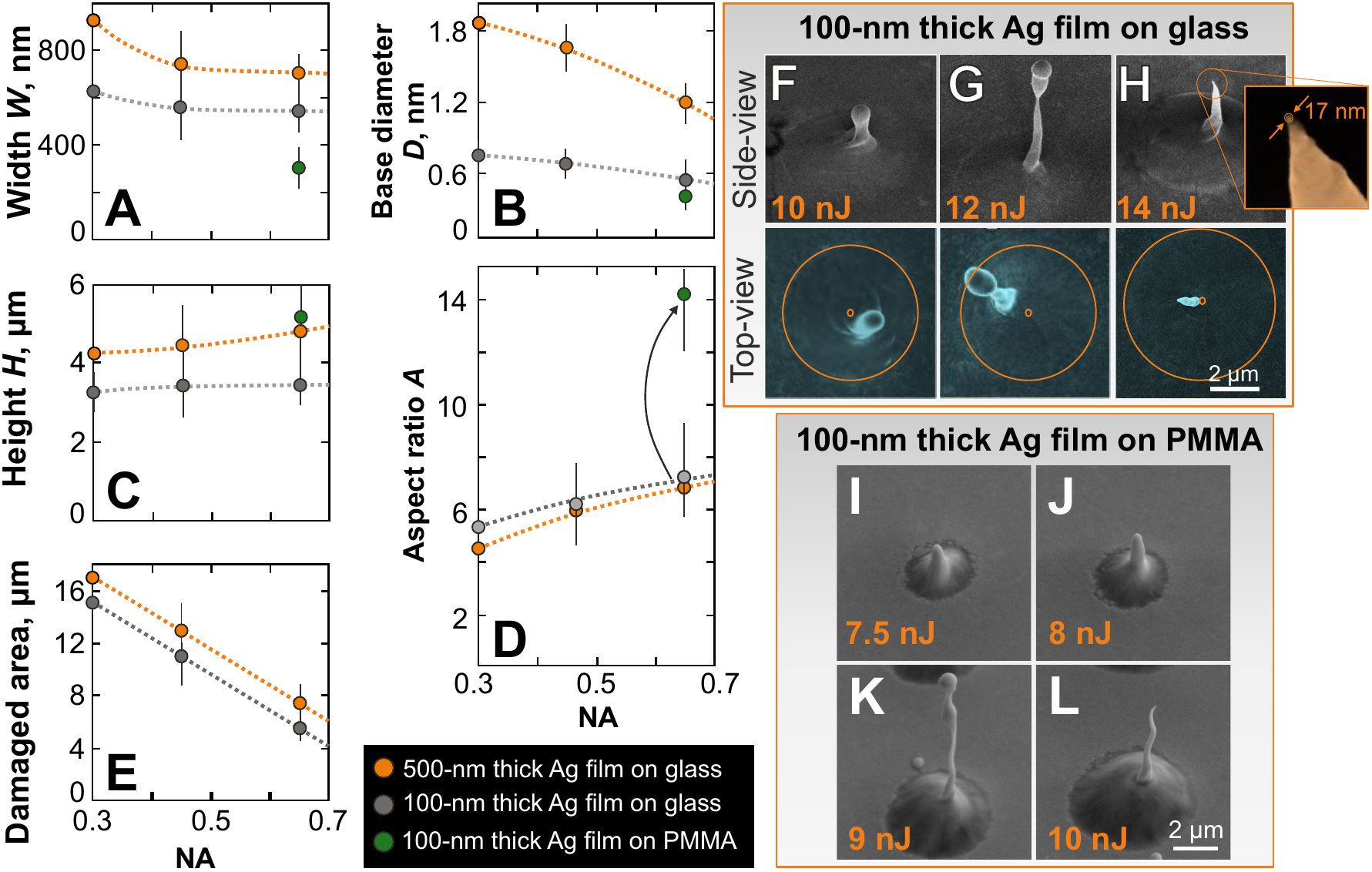}
\caption{(A-D) Averaged thickness- and NA-dependent width w (A), base diameter D (B), height H (C) and aspect ratio A, of the chiral nanojets produced under single-pulse ablation of the 500- and 100-nm thick Ag films, covering the silica glass (orange and grey circles, respectively) and PMMA (green circles) substrates. (E) Diameter of the damaged area on the surface of the 500-nm Ag film versus NA. (F-H) Side-view (view angle of 40°) and top-view (in false color) SEM image pairs of the twisted nanojets produced under vortex-pulse ablation of the 100-nm thick Ag film, covering the silica glass (F-H) and PMMA (I-L) substrates. The pulse energy in each image row increases from left to right. The vortex pulses were focused by the 0.65-NA objective. Inset in the figure (H) shows the magnified view of the nanojet tip.}
\end{figure*}
\section{\label{sec:level3}Results and discussions}
Series of energy-resolved SEM images (Fig. 2(A-D)) illustrates the formation and evolution of the twisted nanojets produced under single-pulse irradiation of the 500-nm thick Ag film, covering the silica glass substrate, with ns-laser vortex pulse focused at NA=0.3. Besides the evident twisted shape of the produced nanojets with their clockwise rotational symmetry (the sign of vortex helicity), expectedly coinciding with the vortex pulse handedness, their formation process appears to be very similar to that for common nanojets fabricated under single-pulse ablation of noble (semi-noble) metal films with Gaussian-shape (zero-OAM) ns-laser pulses [21-26]. The main energy-resolved steps include (i) accumulation of the molten material at the spatial center of the beam through thermocapillary forces and melt flows, which rapidly thin the peripheral part, (ii) formation of the liquid nanojet, which undergoes a Rayleigh-Plateau hydrodynamic instability, resulting in appearance and ejection of the molten droplets and, finally, (iii) formation of the through hole by breaking the significantly thinned area, surrounding the nanojet, at increased pulse energy (this stage is not shown in the SEM images). The tipped twisted nanojets are expectedly fabricated for typical pulse energies sufficient to trigger the ejection of all molten droplets from the resolidifying tip. Under such condition, the produced twisted nanojet can be characterized by very fine tips with the typical averaged curvature radius Rtip~12±6 nm, as it was shown by the high-resolution SEM imaging (see inset in Fig. 2(C)).\\
\begin{figure*}
\includegraphics[width=0.75\textwidth]{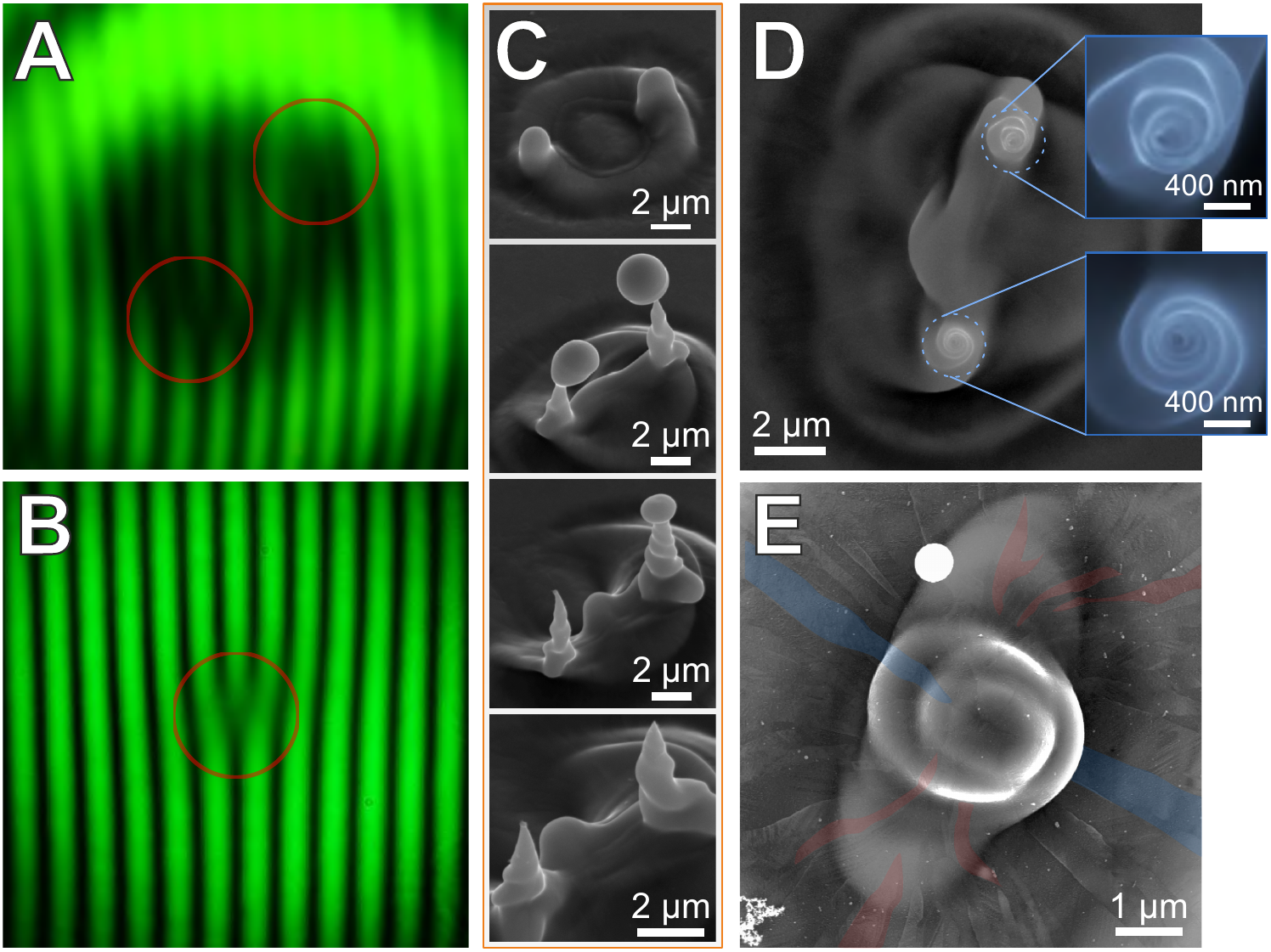}
\caption{Fringe patterns of the interfering plane wave and its donut-shaped replica generated by S-waveplate for the circular (A) and elliptical (B) polarizations. The circles indicate the phase discontinuities. (C) Side-view (view angle of 40°) SEM images of doubled twisted silver nanojets produced under single-pulse ablation of 500-nm thick Ag film with donut-shaped pulses carrying two vortices with the opposite signs of vortex helicity. (D) Top-view SEM image of similar double-jet structure with the magnified images of both nanojets given in the insets. The vortex pulses were focused by the 0.3-NA objective. (E) Top-view high-resolution SEM image, showing the package of nanocrystallites within the single chiral nanojet. Several nanocrystallites were highlighted by false colors to shows the “nanocrystallite bending” effect.}
\end{figure*}
\indent For more tightly focused vortex pulses, similar chiral nanojets, having ultrafine tips, can be produced on the surface of the 500-nm thick Ag film under the single-pulse ablation (see Fig. 2(E-L)). The main difference, evident from the analysis of the presented SEM images, consists in the amount of the molten material involved in the formation process, yielding in variation of the main geometrical parameters of the produced twisted conically shaped nanojets – height (H), width (w), base diameter D and height-to-width (aspect) ratio A. By varying the focusing conditions and systematically studying the formation process of the nanojets under single-pulse ablation of such films, we identify the several general trends characterizing the effect of the objective’s NA. First, the formation of the thinner chiral jets (simultaneous decrease of w and D), providing evident increase of their aspect ratio A (Fig. 3(A-D), is observed for the more tightly focused vortex pulses. With the typical width w of the silver nanojets being varied between 0.95 and 0.68 μm for the tested NA range (almost 4-fold smaller values, comparing to those reported for nanojets produced on the bulk tantalum surface [10]), the aspect ratio A increases two-fold, in its turn (orange circles in Fig. 3D). Second, the average height H accessed for the tipped nanojets without the droplets atop was found to be slightly affected by the NA value (Fig. 3(C)). Similar observations were reported in [10]. Finally, the chirality of the nanojets decreases versus NA. In more general words, the chiral nanojets produced under high-NA focusing (NA=0.65) become less twisted (averaged number of the turnovers per single nanojet decreases) and more similar to those produced under zero-OAM Gaussian pulse ablation [21-26]. In particular, this can indicate that the formation process is directly connected with the resolidification time, which is expectedly smaller for the structures produced under the tight focusing conditions.\\
\indent Similar NA-dependent trends were observed for the chiral nanojets produced on the surface of the 100-nm thick Ag film, covering the glass substrate (grey circles in the Fig. 3(A-D)). Meanwhile, excluding the aspect ratio value A, all other geometric parameters of such nanojets demonstrate evident decrease, comparing to those measured for nanojets produced on the 500-nm thick Ag film, apparently owing to smaller amount of the involved molten material. The SEM study of the nanojets produced on the 100-nm thick Ag film at tight focusing (NA=0.65) indicates their chiral shape as well as pronounced shift of the nanojet position, comparing to the center of both the optical spot and the damaged area (marked by the orange circles in Fig. 3(F-H)), indicating the rotational motion in the direction coincided with the vortex handedness. Also, regardless of the tested experimental parameters (either the objective NA or the film thickness), the average curvature radius Rtip of the nanojet tip was found to be smaller than 17 nm in all cases.\\
\indent In terms of vortex pulse printing of twisted nanojet arrays, it is also important to evaluate their ultimate package density, measuring the size of the film damaged area. Fig. 3(E) shows the typical lateral size of the twisted nanojets as a function of the objective NA, demonstrating the linear scalability and potential few-micron package density at high-NA focusing, which will be studied in details in our forthcoming papers. Finally, since the thickness of the metal film in the microbump area is very small, its exposure to the second laser pulse usually results in destruction of the microbump and appearance of a through hole. In this respect, it seems impossible to use multi-pulse irradiation to reshape or enhance the chirality of the produced jets [10]. Meanwhile, as it was shown in this paper, the large set of geometrical parameters can be realized for nanojets via tuning corresponding focusing conditions and film thicknesses.\\
\indent One additional experimental aspect, uncovering formation and enabling extra tunability of nanojet shapes, which will be only briefly addressed in this paper, consists in variation of supporting substrate, which determines adhesion and acoustic impedance matching between the film and substrate. The “substrate effect” was studied by comparing typical geometrical parameters of the twisted nanojets produced on the surface of 100-nm thick Ag films on the silica glass and PMMA substrates at different pulse energies. With almost twice lower ablation threshold measured for the case of PMMA substrate, using the common linear approximation of the squared diameters of through holes D2 versus natural logarithm of applied pulse energy ln(E) [27], the typical width of the nanojets decreases considerably (Fig.3(A)), yielding in averaged 2.2-fold increase in the aspect ratio A values (see Fig. 3(D)). Similar “substrate effect” was found for the ordinary non-chiral nanojets produced under ablation of the same Ag film with zero-OAM pulses, pointing out the adhesion as one of the key factors, affecting the nanojet formation process and achievable parameter range, in its turn. Also, it should be stressed that ablation experiments undertaken for the 1000-nm thick Ag film, covering the glass substrate, did not demonstrate the formation of regular microscale nanojets for the whole tested NA range, while some rotational movement “fingerprints” were identified from the SEM analysis of the produced surface structures (not shown here).\\
\indent As it was mentioned above, the S-waveplate was used to generate the ns-laser vortex pulses, carrying OAM ℓ=±1, from circularly polarized ones. It is known that under illumination of the S-waveplate with elliptically polarized light, two optical vortices with opposite SAM handedness and opposite OAM handedness can appear [28,29]. This inherent feature of the S-waveplate was used in this paper to generate a donut-shaped beam, carrying two optical vortices with opposite handedness. To do this, for the fixed position of the S-waveplate, we rotated the quarter-wave plate to generate elliptically polarized light, while simultaneously detecting the interference pattern produced by the generated donut-beam and the plane wave from the reference arm of the common Mach-Zehnder interferometer scheme. For a certain position of the quarter-wave plate, the single “fork” in the interference pattern converts into two opposite “forks”, indicating the formation of the vortices with opposite helicity signs (Fig. 4(A,B)) without strong deformations of the donut-shaped intensity distribution.\\
\indent Surprisingly, the ablation of the 500-nm thick Ag film, covering the glass substrate with the donut-beam, carrying two vortices, produces two chiral nanojets with opposite handedness, as it is indicated by the series of the energy-resolved SEM images (Fig. 4(C,D)). At the increasing pulse energy, each separate nanojet undergoes the evolution, similar to those observed for single twisted nanojet, finally evolving into tipped nanoneedle with the pronounced chirality. This remarkable demonstration also indicates that, by tailoring complex intensity and phase distributions via DOEs or other optical elements, the complex patterns with multiple chiral plasmonic nanoneedles can be produced under single-pulse ablation.\\
\indent Finally, the similarity of our present observations and the previous extensive experience in fabrication of thin-film nanojets, using short- and ultrashort laser pulses [24-26,30], provokes us to compare formation of the non- and chiral nanoneedles. In comparison to previously reported results [10,17], where no nanojets were observed under irradiation of the bulk metal target with the zero-OAM Gaussian pulses, for silver films studied in this paper, both Gaussian- and donut-shaped beams produce nanojets, having non- and chiral shapes, respectively. In particular, this indicates that for almost similar general mechanism, underlying the formation of the nanojets and associated with temperature-gradient-driven thermocapillary flow of the molten film, strong rotational movement appears under vortex-pulse irradiation, twisting molten material in the direction, coinciding with the vortex helicity sign. Fingerprints of such rotational movement can be found on both microscale -  in the chiral shape of the resolidified nanojets and in the central symmetry breaking of the nanojet spatial position (Fig. 3(F-H)), and nanoscale – in the bending of some nanocrystallites (Fig. 4(E)), typically having radial-symmetry arrangement [25]. High-resolution SEM imaging of produced twisted nanojets shows that such effect is observed only near the nanojet areas, where twisting thermocapillary flows, strong enough to perturb the recrystallization wave, appear (Fig. 4(E)). The helical thermocapillary melt flows within the evolving nanojet and surrounding microbump area possibly can originate from the corresponding characteristic spiral-like (or more complex, [9]) intensity distribution. The origin of such spiral-shaped intensity distribution and corresponding temperature profile on the metal film surface can be explained in terms of optical interference of the incident donut-shaped beam with the spherical wave reflected/scattered from the evolving surface profile of the molten metal film. As any considerable surface profile evolution of the initially smooth metal film starts after passing the electron-phonon relaxation time (few picosecond for noble metal film), helical shape of the nanojets is expected to disappear for pulse durations shorter, than this time. Similar observation was reported for vortex-beam ablation of silicon target [15]. The detailed picture of appearance of such secondary reflected/scattered wave and its interference with the incident one will be a subject of our ongoing experimental studies and theoretical modeling. We believe that this possible alternative explanation of nanojet helicity, together with the present one [15,17], contributes to the basic understanding and supports new elucidating studies of matter structuring by structured light.

\section{\label{sec:level4}Conclusions}

To conclude, nanosecond vortex pulses generated by passing circularly polarized radiation through S-waveplate were found to produce twisted nanojets under single-pulse ablation of Ag films. Main geometric parameters of the produced chiral nanojets, such as height, width and an aspect ratio, were shown to be tunable in a wide range by varying metal film thickness, supporting substrate type, and the optical size of the vortex beam. Donut-shaped vortex nanosecond laser pulses, carrying two vortices with opposite handedness, were demonstrated to produce two chiral nanojets twisted in opposite directions. The results provide new important insights into fundamental physics of the vectorial laser-beam assisted mass transfer in metal films and demonstrate the great potential of this technique for fast easy-to-implement fabrication of chiral plasmonic nanostructures. We believe the obtained results provide useful and important insights into fundamental picture of the vectorial-beam driven mass transfer in metal films, as well as demonstrate great potential of the direct vortex-beam ablation as a versatile fabrication technique for chiral nanophotonics and plasmonics.

\section*{Funding}

 Authors from IACP and FEFU are grateful for partial support to the Russian Foundation for Basic Research (Projects nos. 14-29-07203 - of\_m, 15-02-03173-a, 17-02-00571-a, 17-02-00936-a) and to FASO through “Far East Program”. Authors from SNRU are grateful for partial support to the Russian Foundation for Basic Research (Project no. 16-29-11698). A.A. Kuchmizhak acknowledges the partial support from RF Ministry of Science and Education (Contract No. МК-3287.2017.2) through the Grant of RF President. S.I. Kudryashov is grateful for the partial support by the Government of the Russian Federation (Grant 074-U01) through ITMO Visiting Professorship Program, and by the Presidium of Russian Academy of sciences. E.V. Pustovalov is grateful for the partial support by the Russian Ministry of Education and Science (grant \# 3.7383.2017).

\section*{References}

\footnotesize{\noindent 1.	N. M. Litchinitser, “Structured light meets structured matter,” Science 337(6098), 1054-1055 (2012).\\
2.	K. Y. Bliokh, F. J. Rodríguez-Fortuño, F. Nori and A. V. Zayats, “Spin-orbit interactions of light,” Nat. Photon. 9(12), 796-808 (2015).\\
3.	J. Kaschke and M. Wegener, “Optical and infrared helical metamaterials,” Nanophotonics 5(4), 510-523 (2016).\\
4.	G. Rui and Q. Zhan, “Tailoring optical complex fields with nano-metallic surfaces,” Nanophotonics 4(1), 2-25 (2015).\\
5.	A. Gopal, M. Hifsudheen, S. Furumi, M. Takeuchi and A. Ajayaghosh, “Thermally assisted photonic inversion of supramolecular handedness,” Angew. Chem. 124(42), 10657-10661 (2012).\\
6.	J. Kumar, T. Nakashima and T. Kawai, “Circularly polarized luminescence in chiral molecules and supramolecular assemblies,” J. Phys. Chem. Lett. 6(17), 3445-3452 (2015).\\
7.	H. Li, J. Cheng, Y.  Zhao, J. W. Lam, K. S. Wong, H. Wu and B. Z. Tang, “L-Valine methyl ester-containing tetraphenylethene: aggregation-induced emission, aggregation-induced circular dichroism, circularly polarized luminescence, and helical self-assembly,” Mater. Horiz. 1(5), 518-521 (2014).\\
8.	T. Ikeda, T. Masuda, T. Hirao, J.Yuasa, H. Tsumatori, T. Kawai and T.Haino, “Circular dichroism and circularly polarized luminescence triggered by self-assembly of tris(phenylisoxazolyl)- benzenes possessing a perylenebisimide moiety,” Chem. Commun. 48, 6025−6027 (2012).\\
9.	J. Yeom, B. Yeom, H. Chan, K. W. Smith, S. Dominguez-Medina, J. H. Bahng, G. Zhao, W.-S. Chang, S,-J. Chang, A. Chuvilin, D. Melnikau, A.L. Rogach, P. Zhang, S. Link, P. Kral, N.A. Kotov, “Chiral templating of self-assembling nanostructures by circularly polarized light,” Nat. Mat. 14(1), 66-72 (2015).\\
10.	K. Toyoda, K. Miyamoto, N. Aoki, R. Morita and T. Omatsu, “Using optical vortex to control the chirality of twisted metal nanostructures,” Nano Lett. 12, 364-3649 (2012).\\
11.	T. Omatsu, K. Chujo, Miyamoto, M. Okida, K. Nakamura, N. Aoki and R. Morita, “Metal microneedle fabrication using twisted light with spin,” Opt. Express 18(17), 17967-17973 (2010).\\
12.	A. Ambrosio, L. Marrucci, F. Borbone, A. Roviello, and P. Maddalena, “Light-induced spiral mass transport in azo-polymer films under vortex-beam illumination,” Nat. Comm. 3, 989 (2012).\\
13.	A. Kravchenko, A. Shevchenko, V. Ovchinnikov, A. Priimagi and M. Kaivola,”Optical interference lithography using azobenzene-functionalized polymers for micro- and nanopatterning of silicon,”.Adv. Mater. 23, 4174–4177 (2011).\\
14.	M. Watabe, G. Juman, K. Miyamoto and T. Omatsu, “Light induced conch-shaped relief in an azo-polymer film,” Sci. Rep. 4, 4281 (2014).\\
15.	F. Takahashi, K. Miyamoto, H. Hidai, K. Yamane, R. Morita and T. Omatsu, “Picosecond optical vortex pulse illumination forms a monocrystalline silicon needle,” Sci. Rep 6, 21738 (2016).\\
16.	J. J. Nivas, H.Shutong, K. K. Anoop, A. Rubano, R. Fittipaldi, A. Vecchione, D. Paparo, L. Marrucci, R. Bruzzese and S. Amoruso, “Laser ablation of silicon induced by a femtosecond optical vortex beam,” Opt. Lett. 40(20), 4611-4614 (2015).\\
17.	K. Toyoda, F. Takahashi, S. Takizawa, Y. Tokizane, K. Miyamoto, R. Morita and T. Omatsu, “Transfer of light helicity to nanostructures,” Phys. Rev. Lett. 110(14), 143603 (2013).\\
18.	C. Hnatovsky, V. G.Shvedov, N. Shostka, A.V. Rode and W. Krolikowski, “Polarization-dependent ablation of silicon using tightly focused femtosecond laser vortex pulses,” Opt. Lett. 37(2), 226-228(2012).\\
19.	C. Hnatovsky, V.Shvedov, W. Krolikowski and A. Rode, “Revealing local field structure of focused ultrashort pulses,” Phys. Rev. Lett. 106(12), 123901 (2011).\\
20.	M. Beresna, M. Gecevičius, P. G. Kazansky and T. Gertus, “Radially polarized optical vortex converter created by femtosecond laser nanostructuring of glass,” Appl. Phys. Lett. 98, 201101 (2011).\\
21.	Y. Nakata, N. Miyanaga, K. Momoo, T. Hiromoto, “Solid–liquid–solid process for forming free-standing gold nanowhisker superlattice by interfering femtosecond laser irradiation,” Appl. Surf. Sci. 274, 27-32 (2013).\\
22.	C. Unger, J. Koch, L. Overmeyer, B.N. Chichkov, “Time-resolved studies of femtosecond-laser induced melt dynamics,” Opt. Express 20, 24864-24872 (2012).\\
23.	A. Kuchmizhak, S. Gurbatov, Y. Kulchin, O. Vitrik, “Fabrication of porous metal nanoparticles and microbumps by means of nanosecond laser pulses focused through the fiber microaxicon,” Opt. Express 22(16), 19149-19155 (2014).\\
24.	J. P. Moening, S.S. Thanawala, D. G. Georgiev, “Formation of high-aspect-ratio protrusions on gold films by localized pulsed laser irradiation,” Appl. Phys. A 95(3), 635-638 (2009).\\
25.	P.A. Danilov, D.A. Zayarny, A.A. Ionin, S.I. Kudryashov, T.T.H. Nguyen, A. Rudenko, I. N. Saraeva, A. A. Kuchmizhak, O. B. Vitrik, Yu. N. Kulchin, “Structure and laser-fabrication mechanisms of microcones on silver films of variable thickness,” JETP Lett. 103(8), 549-552 (2016).\\
26.	A. Kuchmizhak, S. Gurbatov, O. Vitrik, Y. Kulchin, V. Milichko, S. Makarov, S. Kudryashov, “Ion-beam assisted laser fabrication of sensing plasmonic nanostructures,” Sci. Rep. 6, 19410 (2016).\\
27.	J.M. Liu, “Simple technique for measurements of pulsed Gaussian-beam spot sizes,” Opt. Lett. 7, 196 -198 (1982).\\
28.	M. Beresna, M. Gecevicius, and P. G. Kazansky, “Polarization sensitive elements fabricated by femtosecond laser nanostructuring of glass,” Opt. Mater. Express 1, 783–795 (2011).\\
29.	M. Gecevicius, R. Drevinskas, M. Beresna, and P.G. Kazansky, “Single beam optical vortex tweezers with tunable orbital angular momentum,” Appl. Phys. Lett. 104, 231110 (2014).\\
30.	A. Kuchmizhak, S. Gurbatov, A. Nepomniaschiy, A. Mayor, Y. Kulchin, O. Vitrik, S. Makarov, S. Kudryashov, A. Ionin, “Hydrodynamic instabilities of thin Au/Pd alloy film induced by tightly focused femtosecond laser pulses,” Appl. Surf. Sci. 337, 224-229 (2015).}
\end{document}